\documentclass[aps,prd,twocolumn,showpacs,superscriptaddress,groupedaddress]{revtex4-1} 
\usepackage{graphicx}  
\usepackage{dcolumn}   
\usepackage{bm}        
\usepackage{amssymb}   

\usepackage{booktabs}
\usepackage{longtable}
\usepackage{overpic}
\usepackage{multirow}
\usepackage{amsmath}
\usepackage{color}
\usepackage{xspace}
\usepackage{hyperref}

\newsavebox{\tablebox}

\def\BES {\text{BES\uppercase\expandafter{\romannumeral3}}}
\def\BTWO {\text{Belle{\ }\uppercase\expandafter{\romannumeral2}}}

\newcommand{\gev}{~\rm GeV}

\newcommand{\ee}{e^{+}e^{-}}
\newcommand{\eeto}{e^{+}e^{-} \to }
\newcommand{\deltapp}{\Delta^{++} \bar{\Delta}^{--}}

\newcommand{\deltao}{\Delta^{0} \bar{\Delta}^{0}}

\newcommand{\deltappdecay}{\Delta^{++} \to p \pi^{+}}

\newcommand{\pppipi}{p \bar{p} \pi^{+}\pi^{-}}

\newcommand{\deltappany}{\Delta^{++} \bar{p}\pi^{-}}
\newcommand{\anydeltapp}{p\pi^{+} \bar{\Delta}^{--}}

\newcommand{\lambdapair}{\Lambda \bar{\Lambda}}

\begin{document}


\title{\boldmath Production of doubly-charged $\Delta$ baryon in $\ee$ annihilation at energies from $2.3094$ to $2.6464$~GeV}
\author{
M.~Ablikim$^{1}$, M.~N.~Achasov$^{12,b}$, P.~Adlarson$^{72}$, R.~Aliberti$^{33}$, A.~Amoroso$^{71A,71C}$, M.~R.~An$^{37}$, Q.~An$^{68,55}$, Y.~Bai$^{54}$, O.~Bakina$^{34}$, R.~Baldini Ferroli$^{27A}$
% ...
\\
\vspace{0.2cm}
(BESIII Collaboration)\\
\vspace{0.2cm} {\it
$^{1}$ Institute of High Energy Physics, Beijing 100049, People's Republic of China\\
% ...
$^{12}$ G.I. Budker Institute of Nuclear Physics SB RAS (BINP), Novosibirsk 630090, Russia\\
% ...
$^{27}$ INFN Laboratori Nazionali di Frascati , (A)INFN Laboratori Nazionali di Frascati, I-00044, Frascati, Italy; (B)INFN Sezione di  Perugia, I-06100, Perugia, Italy; (C)University of Perugia, I-06100, Perugia, Italy\\
% ...
$^{33}$ Johannes Gutenberg University of Mainz, Johann-Joachim-Becher-Weg 45, D-55099 Mainz, Germany\\
$^{34}$ Joint Institute for Nuclear Research, 141980 Dubna, Moscow region, Russia\\
% ...
$^{37}$ Liaoning Normal University, Dalian 116029, People's Republic of China\\
% ...
$^{54}$ Southeast University, Nanjing 211100, People's Republic of China\\
$^{55}$ State Key Laboratory of Particle Detection and Electronics, Beijing 100049, Hefei 230026, People's Republic of China\\
% ...
$^{68}$ University of Science and Technology of China, Hefei 230026, People's Republic of China\\
% ...
$^{71}$ University of Turin and INFN, (A)University of Turin, I-10125, Turin, Italy; (B)University of Eastern Piedmont, I-15121, Alessandria, Italy; (C)INFN, I-10125, Turin, Italy\\
$^{72}$ Uppsala University, Box 516, SE-75120 Uppsala, Sweden\\
% ...
$^{b}$ Also at the Novosibirsk State University, Novosibirsk, 630090, Russia\\
% ...
}
}

\date{\today}

\begin{abstract}
The processes $\eeto \deltapp$ and $\eeto\Delta^{++} \bar{p} \pi^{-} + c.c.$ are studied for the first time with $179~{\rm pb}^{-1}$ of $\ee$ annihilation data collected with the BESIII detector at center-of-mass energies from $2.3094\gev$ to $2.6464\gev$. 
No significant signal for the $\ee\to\deltapp$ process is observed and the upper limit of the Born cross section is estimated at each energy point.
For the process $\eeto\deltappany + c.c.$, a significant signal is observed at center-of-mass energies near 2.6454~GeV and the corresponding Born cross section is reported.
\end{abstract}

\maketitle

\section{Introduction}
Baryons, specifically the proton and neutron, are the basic building blocks 
of matter.  These half-integer spin fermions are comprised of three valence 
quarks bound together by the strong interaction.  
The most common baryons are the ones from the spin 1/2 SU(3) octet, whose properties have been extensively studied in electron-positron collision experiments~\cite{baryons}. 
The cross-section lineshape of many baryon pair production processes seem to display the common feature of a plateau starting from threshold, including $p\bar{p}$~\cite{pp}, $n\bar{n}$~\cite{nn}, $\Lambda\bar{\Lambda}$~\cite{Lambda}, $\Sigma\bar{\Sigma}$~\cite{Sigma}, $\Xi\bar{\Xi}$~\cite{Xi}, and $\Lambda_{c}^{+}\bar{\Lambda}_c^{-}$~\cite{Lambdac}. 

However, the baryon decuplet, which is comprised of the spin $3/2$ SU(3) baryons, is not fully investigated in electron-positron collisions. 
Compared to the baryon octet, the decuplet is a simpler system as the wave functions of the decuplet baryons are symmetric under flavor exchange. 
The lightest member of the decuplet is the $\Delta$ baryon 
with a mass of 1232~MeV/$c^2$, which is heavier than the nucleon by about 300~MeV/$c^2$.
Experimental information on $\Delta$ baryons comes mostly from experiments in the space-like region~\cite{DeltaSpacelike}, but rarely from the time-like region.
The $\ee$ annihilation process can provide information on time-like form factors of the $\Delta$, similar to the practice for octet baryons~\cite{baryons}. 

In a naive perturbative description of the $\ee$ annihilation into baryons, 
the virtual time-like photon is first coupled to a primary $q\bar{q}$ pair, which then hadronizes by popping two additional quark-antiquark pairs from the vacuum. 
The total perturbative cross section $\sigma$ at a given center-of-mass (c.m.) energy $\sqrt{s}$ is obtained by superposing the amplitudes with different flavors $q$ in the primary $q\bar{q}$ pair and squaring the result,
$\sigma(\eeto~N\bar{N})~\propto~|\sum Q_{q}a_{q}^N(s)|^2$, 
%
where $a_{q}^N$ represents the amplitude of producing the baryon $N$ with a given primary flavor $q$ 
and $Q_q$ denotes the charge of flavor $q$. 
These amplitudes are determined by the baryon wave functions \cite{nnpuzzle}.
Since $\Delta$ baryons have totally symmetric wave functions, the corresponding amplitudes are equal, $a_u^\Delta = a_d^\Delta  \equiv a^\Delta$.
For all 4 members of the $\Delta$ multiplet, $\Delta^{++}, \Delta^{+}, \Delta^{0}, \Delta^{-}$, their relative yields predicted by perturbative theory are 
$\Delta^{++}:\Delta^{+}:\Delta^0:\Delta^- = 4:1:0:1$
\cite{nnpuzzle,qcdbaryonpair}.
However, the relative yield within the multiplet can also be obtained by the relevant Clebsch-Gordan (C-G) coefficients~\cite{PDG} and corrected by the corresponding reduced matrix elements for isospin 1 and 0. 
For $\Delta$ baryons, one expects the relative yields 
$\sigma(\Delta^{++}) = \sigma(\Delta^{-})$ and $
\sigma(\Delta^{+})=\sigma(\Delta^{0})$ 
from the C-G decomposition even without the precise values of the reduced matrix elements~\cite{nnpuzzle}.
This contradicts the perturbative prediction.  

A more detailed discussion of baryon pair production is carried out by J.~G.~K\"orner and M.~Kuroda in Ref.~\cite{eeBBDesy}, including the baryon decuplet.
Since the $\Delta$ states are spin 3/2 baryons, four form factors are required to fully describe their structure. 
In the framework of the generalized vector-dominance model (GVDM), the form factors can be predicted by assuming they arise from the coupling of many vector mesons in the form of a product of poles. 
Within the GVDM, the $\eeto\deltapp$ process has the largest cross section in the baryon decuplet due to the relatively light mass and double charge of the $\Delta^{++}$ baryon, which can reach tens of pb near the threshold. 

The energy threshold of $\Delta^{++}$ pair production is accessible at BESIII and datasets at different energies have been collected to study baryon pair production. 
The predicted cross section for $\eeto\deltapp$ may be accessible at BESIII, providing a first step toward extraction of the production ratios within the $\Delta$ multiplet in order to test theoretical predictions.  
In this paper, we present a search for the process $\eeto\deltapp$ with the subsequent decay $\deltappdecay$ based on data samples collected with the BESIII detector at 6 c.m.~energies from 2.3094
to 2.6464 GeV, corresponding to a total integrated luminosity of 179 pb$^{-1}$~\cite{Trk, Lum}, as listed in Table~\ref{tab:data}. 
Simultaneously, the single $\Delta$ processes, $\eeto\deltappany$ and its charge conjugate (c.c.), are also studied. 

\begin{table}[htbp]
\caption{Data set and expected cross section. The symbol $\sqrt{s}$ is the c.m.~energy. $\mathcal{L}$ is the integrated luminosity. $\sigma^{\rm theory}$ is the cross section of the $\eeto\deltapp$ process predicted in Ref.~\cite{eeBBDesy}.}
\label{tab:data}
\centering
\renewcommand\arraystretch{1.3}
\setlength{\tabcolsep}{12pt}
	\begin{tabular}{c c c }
\hline
\hline
$\sqrt{s}$ (GeV) & $\mathcal{L}$ (pb$^{-1}$) & $\sigma^{\rm theory}$ (pb)  \\
\hline
$2.3094$ & $21.1$ &  0.0 \\
$2.3864$ & $22.5$ &  0.0\\
$2.3960$ & $66.9$ &  0.0 \\
$2.5000$ & $1.10$ &  19.7 \\
$2.6444$ & $33.6$ & 5.7 \\
$2.6464$ & $34.1$ & 5.6 \\
\hline
\hline
	\end{tabular}
\end{table}

\section{Detector and Data Samples}
The BESIII detector~\cite{Ablikim:2009aa} records symmetric $\ee$ collisions provided by the BEPCII storage ring~\cite{Yu:IPAC2016-TUYA01}, 
which 
operates in the c.m.~energy range from 2.0 to 4.95 GeV, with a peak luminosity of $1\times 10^{33}~{\rm cm}^{-2}{\rm s}^{-1}$ achieved at $\sqrt{s} = 3.77$~GeV.
 BESIII has collected large data samples in this energy region~\cite{besIIIdata}. 
The cylindrical core of the BESIII detector covers 93\% of the full solid angle and consists of a helium-based multilayer drift chamber (MDC), 
a plastic scintillator time-of-flight system (TOF), and a CsI(Tl) electromagnetic calorimeter (EMC), which are all enclosed in a superconducting solenoidal magnet providing a 1.0~T magnetic field~\cite{geometry}. 
The solenoid is supported by an octagonal flux-return yoke with resistive plate counter muon identification modules interleaved with steel. 
The charged-particle momentum resolution at 1~GeV/$c$ is 0.5\%, and the specific ionization energy (${\rm d}E/{\rm d}x$) resolution is 6\% for electrons from Bhabha scattering. 
The EMC measures photon energies with a resolution of 2.5\% (5\%) at 1 GeV in the barrel (end cap) region. 
The time resolution in the TOF barrel region is 68 ps, while that in the end cap region is 110~ps.
%

The {\sc geant4}-based~\cite{Geant4} simulation software package {\sc boost}~\cite{BOOST}, which includes the geometric and material description of the BESIII detector and the detector response,  is used to produce Monte Carlo (MC) simulated data samples. 
The initial particles are provided by process-dependent generators, then treated with {\sc boost}. 
Exclusive MC samples are generated using the {\sc ConExc} generator~\cite{ConExc} with initial state radiation (ISR) and vacuum polarization (VP) taken into account to determine the detection efficiencies and to provide shapes
of the involved processes. 
Inclusive hadron production of the type $\eeto {\rm hadrons}$ is simulated by a hybrid generator~\cite{hybrid} to estimate possible background processes and to optimize event selection criteria. 
The beam energy spread of BEPCII is less than 1~MeV at $\sqrt{s} < 3~{\rm GeV}$, which is much smaller than the experimental resolution of the BESIII detector and can be ignored in the simulation.

\section{Event selection}
For the $\eeto\deltapp$ process (with the subsequent decay $\deltappdecay + c.c.$), 
the final state $\pppipi$ is reconstructed for the study of $\Delta$ production.
Candidate events are required to have exactly four reconstructed charged tracks detected in the acceptance of the MDC within a polar angle ($\theta$) range of $|\cos\theta| < 0.93$, where $\theta$ is defined with respect to the $z$-axis, which is the symmetry axis of the MDC. 
For these tracks, the distance of closest approach to the interaction point must be less than 10~cm along the $z$-axis, and less than 1~cm in the transverse plane.
Particle identification (PID) combines measurements of the ${\rm d}E/{\rm d}x$ in the MDC and the flight time in the TOF to calculate a likelihood $\mathcal{L}(h)$ ($h = p, K, \pi$) for each hypothesis of a hadron $h$.
Tracks are identified as protons when the proton hypothesis has the greatest likelihood ($\mathcal{L}(p) > \mathcal{L}(K)$ and $\mathcal{L}(p) > \mathcal{L}(\pi)$).  Charged pions are identified by comparing the pion and kaon hypothesis likelihoods of the remaining tracks, requiring $\mathcal{L}(\pi) > \mathcal{L}(K)$.
Exactly two oppositely charged pions and one proton-antiproton pair are required in each event.  
To improve the momentum and energy resolution and to suppress background events, a four-constraint (4C) kinematic fit imposing four-momentum conservation is performed under the hypothesis $\eeto \pppipi$. 
The $\chi^2$ of the kinematic fit is required to be less than 50. 
The semi-$\Delta$ process, i.e. $\eeto\deltappany$ and its charge conjugate, can be studied simultaneously since it has the same final state particles and similar kinematics. 

\begin{figure}[h]
  \centering
  \includegraphics[width=0.48\textwidth]{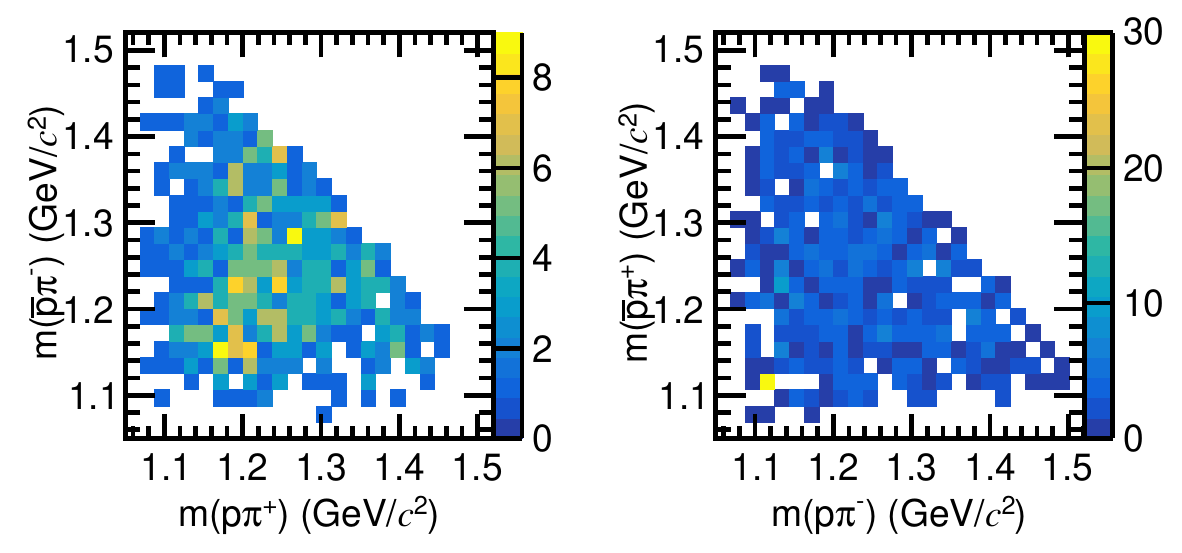}
 \caption{2D distributions of $m(\bar{p}\pi^{-})$ vs.~$m(p\pi^{+})$ (left) and $m(\bar{p}\pi^{+})$ vs.~$m(p\pi^{-})$ (right) for the candidate events selected from data taken at  $\sqrt{s}=2.6444$~GeV.}
 \label{fig:m2d}
\end{figure}

The characteristic signal is expected to appear in the invariant mass spectrum of the $p\pi$ combination, which is denoted as $m(p\pi)$. 
The two-dimensional (2D)  distributions of $m(p\pi^+)$ vs.~$m(\bar{p}\pi^-)$ and $m(p\pi^-)$ vs.~$m(\bar{p}\pi^+)$
of the events selected from data at $\sqrt{s}=2.6444$~GeV are shown in Fig.~\ref{fig:m2d} as an example.  An enhancement around the nominal $\Delta$ baryon mass is visible in the doubly charged combination of $p\pi$ (left plot) while the $\Lambda$ signal is visible in the neutral combination (right plot).
Potential background reactions to the $\eeto\deltapp$ process are studied using inclusive $\eeto {\rm hadrons}$ MC samples. 
Simulated events are subject to the same selection procedure as that applied to the experimental data. 
According to MC simulations, the dominant background stems from $\eeto\deltappany + c.c.,\quad \lambdapair$, and $\pppipi$ processes, which have the same final state particles as the signal reaction. 
Other background channels like $\eeto\rho^0 p\bar{p}$, $\Delta^0 \bar{p} \pi^{+} + c.c.$ and $\deltao$ are also possible but rare, according to the inclusive hybrid MC sample described earlier.

\section{Signal yield and Cross section}
\begin{figure}[htbp]
  \centering
  \includegraphics[width=0.48\textwidth]{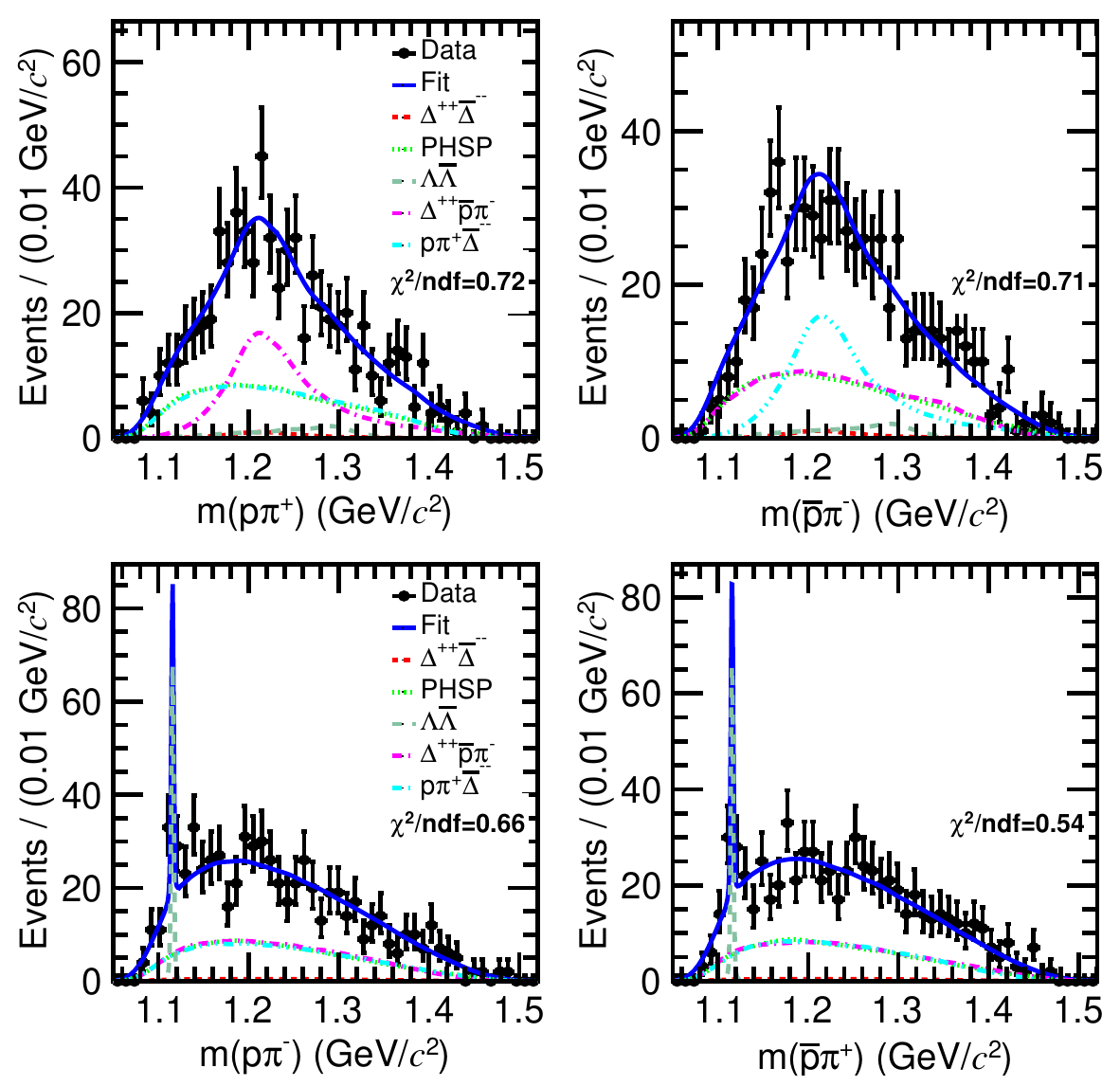}
 \caption{Projections of the simultaneous fit to the 2D distributions of $m(\bar{p}\pi^{-})$ vs.~$m(p\pi^{+})$ and $m(\bar{p}\pi^{+})$ vs.~$m(p\pi^{-})$ of the candidate events selected from data taken at $\sqrt{s}=2.6444$~GeV. The black dots with error bars are data. The blue solid curve is the total fit result. The dashed and dotted curves show the components as indicated in the legends, and $\chi^2/{\rm ndf}$ is the ratio of the $\chi^2$ value to the number of degrees of freedom in the fit.
 }
 \label{fig:fit26444}
\end{figure}

Since the data samples in both 2D distributions are identical, there is a correlation between them, 
which can potentially impact the statistical uncertainty in signal extraction. 
The extent of this influence is contingent upon the correlation of the shapes exhibited by different variable distributions. 
In the case of the $\pppipi$ final state, the variation in spectrum shapes among the different components is not significant in spectra other than the characteristic spectrum due to the final state being a 4-body system with numerous degrees of freedom. 
Consequently, the presence of the correlation does not significantly affect the statistical uncertainty. 
%
Nonetheless, the correlation can help to constrain the yields of the components in a 2D simultaneous fit. 
To determine the signal yield of the $\eeto\deltapp$ process, a simultaneous unbinned maximum likelihood fit is performed to the 2D distributions of $m(\bar{p} \pi^-)$ vs.~$m(p\pi^+)$ and $m(\bar{p}\pi^+)$ vs.~$m(p\pi^-)$. 
This procedure is repeated at each energy point. 
In the fit, the yields of corresponding components in both 2D spectra are obtained with shared parameters.
The signal is modeled with the MC-derived shape of $m(\bar{p} \pi^-)$ vs.~$m(p\pi^+)$ and $m(\bar{p}\pi^+)$ vs.~$m(p\pi^-)$ for the $\eeto\deltapp$ process. 
The background is described by the MC-simulated shape of four dominant components, 
which are phase space (PHSP) \mbox{($\eeto\pppipi$)}, 
semi-$\Delta$ ($\eeto\deltappany$ and $\anydeltapp$, also with the phase space model), 
and \mbox{$\eeto\lambdapair$}, 
while other rarer components are neglected since their distributions are similar to those of the phase space. 
In the analysis, the interference among different components is neglected, 
considering the spin of the combination of a proton and a pion ($p\pi$) tends to be 1/2 when produced at low energy, whereas it is 3/2 in the case of $\Delta$ decay. 
As a result, different components tend to be incoherently produced.
%
Figure~\ref{fig:fit26444} shows the projections of $m(p\pi^+)$, $m(\bar p\pi^-)$, $m(p\pi^-)$ and $m(\bar{p}\pi^+)$ for the simultaneous fit to data taken at at 2.6444 GeV; these indicate that the data is well-described by these five components. 
In extraction of the $\deltapp$ signal, the yields of the semi-$\Delta$ processes, i.e.~$\eeto\deltappany +c.c.$, is obtained simultaneously. 
Although the events are duplicately used in the two 2D distributions, 
the statistical uncertainties of the yields of $\Delta^{++}$ involved processes in the simultaneous fit remain unaffected. 
This is because the target signal, $\Delta^{++}$, is a broad resonance resulting in a negligible discrepancy between the signal and the PHSP in the invariant mass spectra other than the characteristic one, as illustrated in Fig.~\ref{fig:fit26444}. 
The yields are listed in Tables~\ref{tab:fullrecsys} and \ref{tab:fullrecdeltappanyccsys}.
Due to the closeness of the energies, the data samples at 2.6444~GeV and 2.6464~GeV are combined. 

The statistical significance of the signal is determined by comparing the change of the negative log-likelihood value in the fit without the signal contribution and considering the change of the number of degrees of freedom.
The statistical significances are listed in Tables~\ref{tab:fullrecsys} and \ref{tab:fullrecdeltappanyccsys}.
The $\deltapp$ signal is not significant for any of the c.m.~energies.  
A significant semi-$\Delta$ component is observed only above 2.6 GeV.  

\begin{figure}[htbp]
  \centering
      \includegraphics[width=0.45\textwidth]{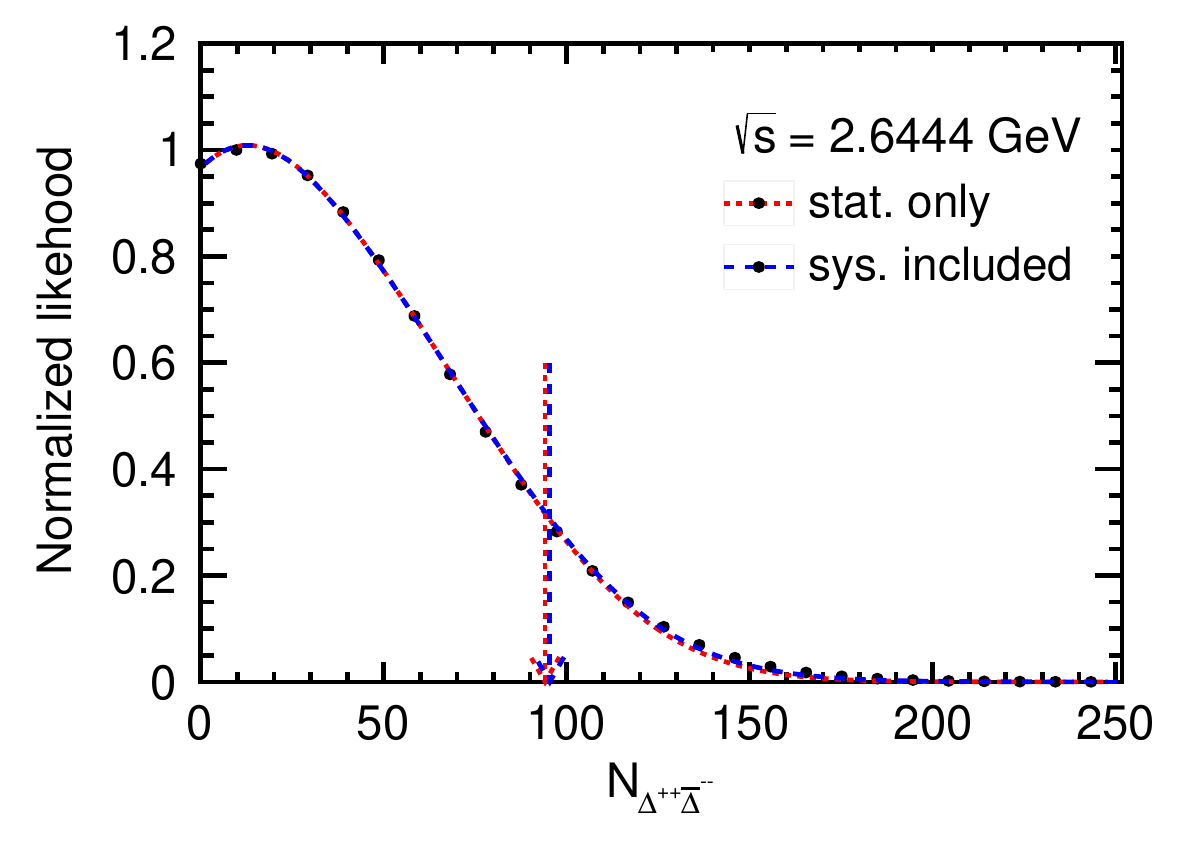}
 \caption{Likelihood distributions vs.~the signal yields of the $\ee\to \deltapp$ process for data taken at \mbox{$\sqrt{s}=2.6444$~GeV}. The red dotted arrow indicates the upper limit at the 90\% C.L. by only considering the statistical uncertainty, while the blue dashed arrow has incorporated the systematic uncertainty.}
 \label{fig:uplimitsys}
\end{figure}

The upper limit on the signal yield of the \mbox{$\eeto\deltapp$} process is determined with a Bayesian approach~\cite{Uplimit} via a likelihood scan. 
The systematic uncertainty is considered by choosing the most conservative result among the fit variations used to evaluate the additive uncertainties and smearing the obtained likelihood curve with a Gaussian function with the width of the multiplicative systematic uncertainty~\cite{UplimitSys}. 
The upper limit on the signal yield at the 90\% confidence level (C.L.), $N_{\rm up}$, is determined by integrating the smeared likelihood function $L(N)$, as illustrated shown in Fig.~\ref{fig:uplimitsys}.  
The upper limit, $N_{\rm up}$, corresponds to 90\% of the full integral, i.e.~ 
$\int_{0}^{N_{\rm up}}L(x){\rm d}x/\int_{0}^{\infty}L(x){\rm d}x = 0.90$.  
The same method is applied to the \mbox{$\eeto\deltappany + c.c.$} process at the energy points with a statistical significance less than $2\sigma$.

\begin{figure}[ht]
  \centering
  \includegraphics[width=0.45\textwidth]{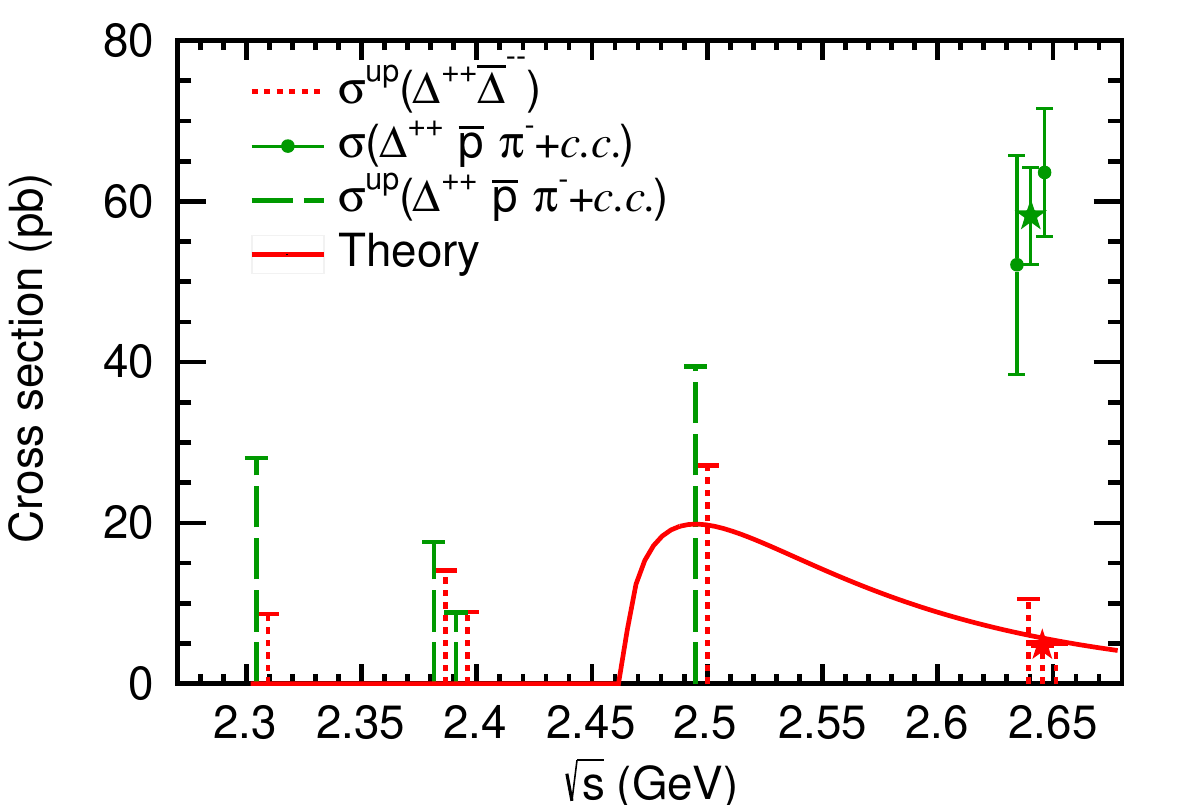}
 \caption{Born cross sections of $\eeto$ $\deltapp$ and $\deltappany+c.c.$. The green dots with solid lines represent the Born cross sections with uncertainties. The red short-dashed and green long-dashed lines show the upper limits. The red solid curve shows theoretical prediction for the $\eeto\deltapp$ process in Ref.~\cite{eeBBDesy}. To avoid the overlap between different processes, the c.m.~energies of the $\eeto\deltappany+c.c.$ process are shifted left by 5 MeV. Pentagrams show combined results of two nearby energy points 2.6444 GeV and 2.6464 GeV, and the results at the two energy points are shifted to left and right by 5 MeV, respectively.}
 \label{fig:crxup}
\end{figure}

The Born cross section is calculated as 

\begin{equation}
\sigma^{B} = \frac{N}{\mathcal{L}\cdot \epsilon \cdot (1+\delta) \cdot Br},
\label{eqcx}
\end{equation}

\noindent
where $N$ is the number of signal events in data, 
$\mathcal{L}$ is the integrated luminosity, 
$\epsilon$ is the detection efficiency, $(1+\delta)$ is the radiative correction factor due to ISR and VP,
and $Br$ is the branching fraction of the decay $\deltappdecay$ (100\%)~\cite{PDG}. 
Both $\epsilon$ and $1+\delta$ are obtained from MC simulation of the signal reaction at each c.m.~energy. 
The input Born cross section lineshape for the $\eeto\deltapp$ process remains constant above the threshold, which is determined by the sum of the masses of internal particles and is equal to 2.464 GeV. 
Below the threshold, the lineshape follows a Breit-Wigner distribution centered at the threshold with a width of 0.165 GeV, representing the width of the $\Delta$ baryon pair. 
The lineshape for the $\eeto\deltappany + c.c.$ process is similar but with the threshold at 2.310 GeV and the width of the Breit-Wigner distribution at 0.117 GeV. 
The results of Born cross sections and upper limits for $\eeto\deltapp$ and $\eeto\deltappany + c.c.$ are reported in Tables~\ref{tab:fullrecsys} and \ref{tab:fullrecdeltappanyccsys}, respectively, and are also shown in Fig.~\ref{fig:crxup}.

\begin{table*}[htbp]
\caption{Born cross sections of $\eeto\deltapp$ at various c.m.~energy points. All symbols are defined the same as those in Eq.~(\ref{eqcx}). 
The $N_{\rm up}$ and $\sigma^{B}_{\rm up}$ are set at the 90\% C.L. for the c.m.~energy points with statistical significance less than $2\sigma$. 
The branching fraction of $\deltapp$ decaying to the final state $\pppipi$ is 100.0\%.
The first uncertainty is statistical and the second one is systematic. The line with an asterisk is a combined result of two nearby energy points $\sqrt{s}=2.6444$ and $2.6464$~GeV.
}
\label{tab:fullrecsys}
\centering
\renewcommand\arraystretch{1.3}
\setlength{\tabcolsep}{12pt}
	\begin{tabular}{c c c c c c c c }
\hline
\hline
$\sqrt{s}$ (GeV) & $N$ ($N_{\rm up}$) & $\epsilon$ & $(1+\delta)$ & $\sigma^B$  ($\sigma^{B}_{\rm up}$) (pb) & Significance ($\sigma$)  \\
\hline
$2.3094$ & $0.0^{+0.3}_{-0.0}$ ($1.4$) & $0.009$ & $0.869$  & $0.0^{+1.9}_{-0.0} \pm 0.0$ ($8.7$) & $0.01$ \\
$2.3864$ & $4.3^{+7.9}_{-4.3}$ ($16.3$) & $0.059$ & $0.864$  & $3.7^{+6.8}_{-3.7} \pm 0.3$ ($14.1$) & $0.56$ \\
$2.3960$ & $17.0^{+12.8}_{-14.1}$ ($35.0$) & $0.068$ & $0.865$  & $4.3^{+3.3}_{-3.6} \pm 0.4$ ($8.9$) & $1.19$ \\
$2.5000$ & $0.0^{+2.3}_{-0.0}$ ($4.7$) & $0.172$ & $0.919$  & $0.0^{+13.2}_{-0.0} \pm 0.0$ ($27.1$) & $0.00$ \\
$2.6444$ & $12.3^{+53.3}_{-12.3}$ ($95.4$) & $0.280$ & $0.962$  & $1.4^{+5.9}_{-1.4} \pm 0.1$ ($10.5$) & $0.24$ \\
$2.6464$ & $0.0^{+11.4}_{-0.0}$ ($46.4$) & $0.282$ & $0.962$  & $0.0^{+1.2}_{-0.0} \pm 0.0$ ($5.0$) & $0.00$ \\
$^*2.6454$ & $0.0^{+26.5}_{-0.0}$ ($87.3$) & $0.281$ & $0.962$  & $0.0^{+1.4}_{-0.0} \pm 0.0$ ($4.8$) & $0.00$ \\
\hline
\hline
	\end{tabular}
\end{table*}

\begin{table*}[htbp]
\caption{Born cross sections of $\eeto\deltappany+c.c.$ at various c.m.~energy points. All symbols are the same as Table~\ref{tab:fullrecsys}. 
}
\label{tab:fullrecdeltappanyccsys}
\centering
\renewcommand\arraystretch{1.3}
\setlength{\tabcolsep}{12pt}
	\begin{tabular}{c c c c c c c c }
\hline
\hline
$\sqrt{s}$ (GeV) & $N$ ($N_{\rm up}$) & $\epsilon$ & $(1+\delta)$ & $\sigma^B$  ($\sigma^{B}_{\rm up}$) (pb) & Significance ($\sigma$)  \\
\hline
$2.3094$ & $0.0^{+0.4}_{-0.0}$ ($1.8$) & $0.004$ & $0.873$  & $0.0^{+6.2}_{-0.0} \pm 0.0$ ($28.1$) & $0.00$ \\
$2.3864$ & $5.2^{+6.9}_{-5.2}$ ($17.0$) & $0.046$ & $0.928$  & $5.4^{+7.2}_{-5.4} \pm 0.5$ ($17.6$) & $0.88$ \\
$2.3960$ & $0.0^{+14.1}_{-0.0}$ ($30.7$) & $0.056$ & $0.932$  & $0.0^{+4.0}_{-0.0} \pm 0.0$ ($8.8$) & $0.01$ \\
$2.5000$ & $0.0^{+5.0}_{-0.0}$ ($6.5$) & $0.156$ & $0.961$  & $0.0^{+30.3}_{-0.0} \pm 0.0$ ($39.4$) & $0.00$ \\
$2.6444$ & $424.8^{+77.6}_{-109.2}$ & $0.247$ & $0.982$ & $52.1^{+9.5}_{-13.4} \pm 4.2$& $3.87$ \\
$2.6464$ & $536.0^{+57.7}_{-58.9}$ & $0.252$ & $0.982$ & $63.6^{+6.8}_{-7.0} \pm 8.0$& $6.26$ \\
$^*2.6454$ & $965.1^{+81.9}_{-82.5}$ & $0.249$ & $0.982$ & $58.2^{+4.9}_{-5.0} \pm 5.9$& $7.09$ \\
\hline
\hline
	\end{tabular}
\end{table*}

\section{Systematic uncertainty}
Various sources of systematic uncertainties concerning the measurement of the Born cross sections are investigated, including event selection criteria, integrated luminosity, radiative correction factor, branching fraction, fit procedure, and the mass and width of the $\Delta^{++}$ baryon.

The uncertainty due to differences between data and MC simulation for the tracking efficiency and PID are investigated using the high-purity control samples of $\eeto K^+K^-\pi^+\pi^-$~\cite{Trk, Trk2} for pions and $\eeto p\bar{p}\pi^+\pi^-$~\cite{ppbar} for protons, 
which are assigned as 1.0\% and 1.0\%, per track, respectively. 
The uncertainty associated with the $\chi^2$ requirement on 4C kinematic fit is also studied by comparing the difference between data and MC simulation. 
A Gaussian function is convolved with the MC-simulated $\chi^2$ shape to better match the data for the $\chi^2$ distribution. 
The change of efficiency with the Gaussian-smeared MC shape is taken as the uncertainty, and ranges from $0.5\%$ to $2.3\%$. 
The integrated luminosity of data at each energy point is measured using large-angle Bhabha events with an uncertainty of 0.8\% following the method in Ref.~\cite{Lum}. 

The uncertainty of the radiative correction factor is obtained by altering the Born cross-section lineshape in the MC simulation.
In the $\deltapp$ case, the nominal lineshape is flat above the threshold with a Breit-Wigner tail below, considering the broad width of the $\Delta$ baryon. 
Although Ref.~\cite{eeBBDesy} predicted a lineshape for the baryon pair production, the shape does not consider the width of the $\Delta$ baryon and becomes 0 below threshold.
To estimate the uncertainty, the shape is replaced by 
a more general power-law shape for a baryon pair production~\cite{pQCDshape}, which is motivated by perturbative quantum chromodynamics and represents the power-law asymptotic behaviour of baryons, convolved with the Breit-Wigner resonance shape of the $\Delta$ baryon. 
In the semi-$\Delta$ case, a three-body PHSP distribution
is chosen as an alternative lineshape to replace the nominal flat distribution. 
The three-body PHSP distribution is an integral of two cascaded quasi-two-body PHSP distributions, i.e. $\sigma \propto \int PS(\sqrt{s}, m_{p\pi}, m_{\Delta}) \cdot PS(m_{p\pi}, m_{p}, m_{\pi}) {\rm d} m^2_{p\pi}$ with $PS(m_a, m_b, m_c)$ to be the two-body PHSP factor of the $a\to b+c$ process~\cite{Phsp}. 
The difference of the product of the efficiency and radiative correction factor after changing the lineshapes is taken as the uncertainty, which is around 1\% at most energy points. 

The uncertainty associated with the branching fraction of $\deltappdecay$ is negligible since this is the only allowed decay mode. 
The uncertainty from the unknown polar angle, $\theta_{\Delta}$, distribution of the $\Delta$ baryon is estimated by dividing the difference of efficiencies between the two extreme angular distributions, $(1+\cos^2\theta_{\Delta})$ and $(1-\cos^2\theta_{\Delta})$, by $\sqrt{12}$ assuming a uniform probability distribution~\cite{UncFlat}, giving uncertainties in the range of $3.8-10.7$\%.

Uncertainties due to the choice of the signal and background shapes are estimated by smearing the shapes with a Gaussian function determined from the difference of the $\Lambda$ peak in data and MC simulation.
The difference on the yields of the $\eeto\deltappany+c.c.$ process in the signal extraction is taken as the uncertainty, and is less than 1\% above 2.6 GeV. 
Since there is almost no signal below 2.6 GeV, the relative systematic uncertainty is not estimated. 
The uncertainty of the $\eeto\deltapp$ process is quoted from that of the $\eeto\deltappany+c.c.$ process. 

Since the $\Delta$ is a wide resonance, the uncertainty due to its mass and width is estimated by using the values in Ref.~\cite{MW} in simulation. 
Specifically, the mass and width are changed from 1.2310 GeV/$c^2$ and 0.1150 GeV, respectively, to 1.2308 GeV/$c^2$ and 0.1109 GeV. 
Then the uncertainty is estimated with the same method for the shapes, and is in the range of $1.5-4.7$\% above 2.6~GeV and 
ignored at lower energies.
The influence of intermediate states is checked for small contributions shown in inclusive hadronic MC samples and several possible resonances.
Among the small contributions, the $\rho^0 p\bar{p}$ component has relatively large influence on the signal yield in the 2D fit. 
The uncertainty from this intermediate process is estimated with the same method for the shapes by comparing the yield of the $\eeto\deltappany+c.c.$ process, which will contribute an uncertainty about 5\%. 
Several possible excited baryons have also been verified and the most likely contribution is found to come from the $N(1710)^+\bar{p}+c.c.$ process with $N(1710)^+\to\Delta^{++}\pi^{-}$. 
Since the $N(1710)^+\bar{p}+c.c.$ channel is part of the $\deltappany+c.c.$ process, the uncertainty is only estimated for the $\deltappany+c.c.$ process above 2.6 GeV and the contribution is at most 5.0\%. 
The intermediate states are correlated and result in reduction of the signal yield.   
A conservative uncertainty is estimated by choosing the largest one among those contributions.
All of the systematic uncertainties of the cross section measurement are summarized in Table~\ref{tab:syssum}.

\begin{table*}[htp]
\caption{Relative systematic uncertainties, in \%, for the Born cross-section measurements. 
Column Trk is the uncertainty from tracking, PID from particle identification, 4C from the 4C kinematic fit, $\mathcal{L}$ from luminosity, $(1+\delta)$ from radiation correction factor, $\alpha$ from angular distribution of baryons, Shape from signal and background shapes in the fit, MW from the mass and width of $\Delta^{++}$ used in simulation, and $R$ from possible intermediate states. For $(1+\delta)$, MW, $R$ and Total, the two numbers before and after ``/'' are for the processes of $\eeto\Delta^{++}\bar{\Delta}^{--}$ and $\Delta^{++}\bar{p}\pi^{-}+c.c.$, respectively.}
\label{tab:syssum}
\begin{center}
\setlength{\tabcolsep}{10.5pt}
\begin{tabular}{c ccc ccc ccc cc}
\hline
\hline
$\sqrt{s}$ (GeV)  & Trk  & PID  & 4C  & $\mathcal{L}$  & $(1+\delta)$ & $\alpha$ & Shape  & MW & $R$ & Total  \\
\hline
$2.3094$  & $4.0$ & $4.0$ & $1.7$ & $0.8$  & $0.5$ / $7.6$  & $10.7$  & $-$  & $-$  & - / -  & $12.3$ / $14.5$ \\
$2.3864$  & $4.0$ & $4.0$ & $1.4$ & $0.8$  & $1.4$ / $1.2$  & $6.2$  & $-$  & $-$  & - / -  & $8.7$ / $8.7$ \\
$2.3960$  & $4.0$ & $4.0$ & $1.5$ & $0.8$  & $1.1$ / $1.2$  & $6.3$  & $-$  & $-$  & - / -  & $8.7$ / $8.7$ \\
$2.5000$  & $4.0$ & $4.0$ & $2.3$ & $0.8$  & $0.8$ / $0.6$  & $3.8$  & $-$  & $-$  & - / -  & $7.3$ / $7.3$ \\
$2.6444$  & $4.0$ & $4.0$ & $0.5$ & $0.8$  & $0.1$ / $0.1$  & $4.8$  & $0.7$  & $1.5$  & $1.2$ / $2.5$  & $7.7$ / $8.0$ \\
$2.6464$  & $4.0$ & $4.0$ & $1.1$ & $0.8$  & $1.4$ / $0.7$  & $4.5$  & $0.3$  & $4.7$  & $9.0$ / $9.0$  & $12.6$ / $12.6$ \\
$^*2.6454$  & $4.0$ & $4.0$ & $0.7$ & $0.8$  & $0.7$ / $0.3$  & $4.6$  & $0.1$  & $4.7$  & $5.1$ / $5.1$  & $10.2$ / $10.1$ \\
\hline
\hline
\end{tabular}
\end{center}
\end{table*}%

\section{Conclusion}
In this paper, the $\eeto\deltapp$ process is searched for with data taken at c.m.~energies from $2.3094$ to $2.6464$~GeV.
There is no significant \mbox{$\ee\to\deltapp$} signal at any energy, while clear signals for the \mbox{$\eeto\deltappany + c.c.$} process are observed above 2.6 GeV. 
Upper limits on the Born cross sections of the $\ee\to \deltapp$ process at the 90\% C.L. are extracted; they are comparable to but lower than the predictions in Ref.~\cite{eeBBDesy}.
With only upper limits, the predicted production ratios among $\Delta$ multiplet members in Ref.~\cite{nnpuzzle} cannot be addressed.  
Higher statistics are required for further studies of $\Delta$ pair production. 

\hspace{0.5 cm}

\section{Acknowledgement}
The BESIII collaboration thanks the staff of BEPCII and the IHEP computing center and the supercomputing center of USTC for their strong support. 
This work is supported in part by National Key R\&D Program of China under Contracts Nos. 
2020YFA0406400, 
2020YFA0406300; 
National Natural Science Foundation of China (NSFC) under Contracts Nos. 
12105276, 
11335008, 11625523, %
11635010, 11705192, 
11735014, %
11835012, 11935015, 11935016, 11935018, 
11950410506, %
11961141012, 12022510, 12025502, 12035009, 12035013, 12061131003, 
12122509, %
12192260, 12192261, 12192262, 12192263, 12192264, 12192265; 
the Chinese Academy of Sciences (CAS) Large-Scale Scientific Facility Program; the CAS Center for Excellence in Particle Physics (CCEPP); 
Joint Large-Scale Scientific Facility Funds of the NSFC and CAS under Contract No. 
U1732263, U1832103, %
U1832207, 
U2032111; %
CAS Key Research Program of Frontier Sciences under Contracts Nos. QYZDJ-SSW-SLH003, QYZDJ-SSW-SLH040; 
100 Talents Program of CAS; The Institute of Nuclear and Particle Physics (INPAC) and Shanghai Key Laboratory for Particle Physics and Cosmology; 
ERC under Contract No. 758462; 
European Union's Horizon 2020 research and innovation programme under Marie Sklodowska-Curie grant agreement under Contract No. 894790; 
German Research Foundation DFG under Contracts Nos. 443159800, 455635585, Collaborative Research Center CRC 1044, FOR5327, GRK 2149; 
Istituto Nazionale di Fisica Nucleare, Italy; Ministry of Development of Turkey under Contract No. DPT2006K-120470; 
National Science and Technology fund; 
National Science Research and Innovation Fund (NSRF) via the Program Management Unit for Human Resources \& Institutional Development, Research and Innovation under Contract No. B16F640076; 
Olle Engkvist Foundation under Contract No. 200-0605; 
STFC (United Kingdom); 
Suranaree University of Technology (SUT), Thailand Science Research and Innovation (TSRI), and National Science Research and Innovation Fund (NSRF) under Contract No. 160355; 
The Royal Society, UK under Contracts Nos. DH140054, DH160214; The Swedish Research Council; 
U. S. Department of Energy under Contract No. DE-FG02-05ER41374.

\end{document}